\documentclass[a4paper,10pt]{revtex4}
\usepackage{graphicx}  
\usepackage{amsmath}   
\usepackage{amssymb}   
\usepackage{bm} 
\usepackage{dcolumn}
\usepackage{color}
\usepackage{mathrsfs}
\usepackage{amsfonts}
\usepackage{varioref}
\usepackage{mathrsfs}
\usepackage{graphicx}
\usepackage{latexsym}
\usepackage{amsmath}
\usepackage{amssymb}
\usepackage{textcomp}
\usepackage{amsbsy}
\usepackage{graphics}
\usepackage{epstopdf}
\usepackage{color}
\usepackage[caption=false]{subfig}

\RequirePackage[colorlinks,citecolor=blue,urlcolor=magenta,linkcolor=blue]{hyperref}
\input epsf

\allowdisplaybreaks[4]

\begin{document}

\tolerance=5000

\title{Microscopic interpretation of generalized entropy}

\author{Shin'ichi~Nojiri$^{1,2}$\,\thanks{nojiri@gravity.phys.nagoya-u.ac.jp},
Sergei~D.~Odintsov$^{3,4}$\,\thanks{odintsov@ieec.uab.es},
Tanmoy~Paul$^{5}$\,\thanks{pul.tnmy9@gmail.com}} \affiliation{
$^{1)}$ Department of Physics, Nagoya University,
Nagoya 464-8602, Japan \\
$^{2)}$ Kobayashi-Maskawa Institute for the Origin of Particles
and the Universe, Nagoya University, Nagoya 464-8602, Japan \\
$^{3)}$ ICREA, Passeig Luis Companys, 23, 08010 Barcelona, Spain\\
$^{4)}$ Institute of Space Sciences (ICE, CSIC) C. Can Magrans s/n, 08193 Barcelona, Spain\\
$^{5)}$ Department of Physics, Visva-Bharati University, Santiniketan -731235, India}


\tolerance=5000

\begin{abstract}
Generalized entropy, that has been recently proposed, puts all the known and apparently different entropies like The Tsallis, the R\'{e}nyi, the Barrow, the Kaniadakis, the Sharma-Mittal and the loop quantum gravity entropy within a single umbrella. However, the microscopic origin of such generalized entropy as well as its relation to thermodynamic system(s) is not clear. In the present work, we will provide a microscopic thermodynamic explanation of generalized entropy(ies) from canonical and grand-canonical ensembles. It turns out that in both the canonical and grand-canonical descriptions, the generalized entropies can be interpreted as the statistical ensemble average of a series of microscopic quantity(ies) given by various powers of $\left(-k\ln{\rho}\right)^n$ (with $n$ being a positive integer and $\rho$ symbolizes the phase space density of the respective ensemble), along with a term representing the fluctuation of Hamiltonian and number of particles of the system under consideration (in case of canonical ensemble, the fluctuation on the particle number vanishes).
\end{abstract}

\maketitle

\section{Introduction}

The growing interest in different entropy functions proposed so far
(like the Bekenstein-Hawking \cite{Bekenstein:1973ur,Hawking:1975vcx}, Tsallis \cite{Tsallis:1987eu}, R\'{e}nyi \cite{Renyi}, Barrow \cite{Barrow:2020tzx}, Sharma-Mittal \cite{SayahianJahromi:2018irq}, Kaniadakis \cite{Kaniadakis:2005zk} and loop quantum gravity entropies \cite{Majhi:2017zao}) towards black hole thermodynamics
as well as towards cosmology \cite{Li:2004rb,Li:2011sd,Wang:2016och,Nojiri:2005pu,Landim:2022jgr,Zhang:2005yz,Elizalde:2005ju,
Ito:2004qi,Gong:2004cb,Khurshudyan:2016gmb,Landim:2015hqa,Li:2008zq,Zhang:2005hs,Li:2009bn,Feng:2007wn,Zhang:2009un,Lu:2009iv,
Micheletti:2009jy,Nojiri:2017opc,Saridakis:2020zol,Barrow:2020kug,Adhikary:2021xym,Nojiri:2019kkp,Nojiri:2020wmh,Komatsu:2023wml,Luciano:2022viz,Lambiase:2023ryq} lead to the natural question that whether there exists a generalized entropy function that can generalize
all these known entropies. The quest becomes even stronger when the entropic cosmology of these entropies proves to be equivalent to the holographic scenario with suitable holographic cut-offs \cite{Nojiri:2021iko,Nojiri:2021jxf}. With these spirits, generalized entropy (with three, four, five and six parameters) has been proposed in several recent works, which gives all the aforementioned known entropies as different representatives at a certain limit of the entropic parameters \cite{Nojiri:2022aof,Nojiri:2022dkr,Odintsov:2022qnn,Odintsov:2023qfj}. In particular, the four- and six-parameter generalized entropies are of the form \cite{Nojiri:2022aof,Nojiri:2022dkr}:
\begin{eqnarray}
 S_\mathrm{4}\left(\alpha_{\pm},\delta,\gamma\right) = \frac{1}{\gamma}\left[\left(1 + \frac{\alpha_+}{\delta}~S\right)^{\delta}
 - \left(1 + \frac{\alpha_-}{\delta}~S\right)^{-\delta}\right]
 \label{intro-1}
 \end{eqnarray}
 and
\begin{eqnarray}
 S_\mathrm{6}\left(\alpha_{\pm},\delta_{\pm},\gamma_{\pm}\right) = \frac{1}{\alpha_+ + \alpha_-}\left[ \left( 1 + \frac{\alpha_+}{\delta_+} \, S^{\gamma_+}
\right)^{\delta_+} - \left( 1 + \frac{\alpha_-}{\delta_-}
\, S^{\gamma_-} \right)^{-\delta_-} \right]\, ,
 \label{intro-2}
\end{eqnarray}
respectively, where $S = A/(4G)$ represents the Bekenstein-Hawking entropy and the argument of the entropy functions contains their respective parameters ($A$ being the area of the apparent horizon in the cosmological scenario). Both of these entropies reduce to all the aforementioned known entropies for suitable limit of the respective parameters, for instance --- $S_\mathrm{4}$ reduces to the Tsallis entropy in the limit of $\alpha_{+} \rightarrow \infty$, $\alpha_{-} = 0$ and $\gamma = \left(\alpha_{+}/\beta\right)^{\beta}$, or regarding the six parameter entropy, $S_\mathrm{6}$ goes to the Tsallis entropy for $\alpha_+ = \alpha_- \rightarrow 0$ and $\gamma_+ = \gamma_-$. Besides the four- and six-parameter generalized entropies, a three-parameter entropy of the form
\begin{eqnarray}
 S_\mathrm{3}\left(\alpha,\delta,\gamma\right) = \frac{1}{\gamma}\left[\left(1 + \frac{\alpha}{\delta}~S\right)^{\delta} - 1\right]
 \label{intro-3}
\end{eqnarray}
has also been proposed in \cite{Nojiri:2022aof}. However, the important point to be noted is that $S_\mathrm{3}$ is not able to generalize $all$ the known entropies, in particular, $S_\mathrm{3}$ cannot be reduced to the Kaniadakis entropy for any limit of the parameters. In this regard, we would like to mention that the four-parameter entropy is the minimal construction of generalized entropy in the sense that the minimum number of parameters required in an entropy function for generalizing all the known entropies is equal to four. The above-mentioned $S_\mathrm{3}$, $S_\mathrm{4}$ and $S_\mathrm{6}$ share the following properties: (a) they obey the third law of thermodynamics, i.e., they vanish in the limit of $S \rightarrow 0$; (b) they show a monotonic behaviour with respect to the variable $S$; (c) they diverge at the point when the Hubble parameter vanishes (i.e $H = 0$) as the Bekenstein-Hawking entropy itself diverges at $H = 0$. Thus the entropy functions $S_\mathrm{3}$, $S_\mathrm{4}$ and $S_\mathrm{6}$ exhibit a time-like singularity at the instance of $H = 0$ which generally occurs at the time of bounce in the context of bouncing cosmology. Here it is important to mention that such diverging behaviour is common to all the known entropies proposed so far, like The Tsallis, the R\'{e}nyi, the Barrow, the Kaniadakis, the Sharma-Mittal and the loop quantum gravity entropy. The issue of such singularity in entropic cosmology gets resolved by \cite{Odintsov:2022qnn} where some of our authors proposed a five-parameter entropy function, which is non-singular and, at the same time, can generalize all the known entropies. The form of the five-parameter non-singular generalized entropy is given by,
\begin{eqnarray}
 S_\mathrm{5}\left(\alpha_{\pm},\delta,\gamma,\epsilon\right) = \frac{1}{\gamma}\bigg[\left\{1 + \frac{1}{\epsilon}\tanh\left(\frac{\epsilon \alpha_+}{\delta}~S\right)\right\}^{\delta}
- \left\{1 + \frac{1}{\epsilon}\tanh\left(\frac{\epsilon \alpha_-}{\delta}~S\right)\right\}^{-\delta}\bigg]
 \label{intro-4}
\end{eqnarray}
which, due to the presence of the $\mathrm{tanh}$ function, remains singular-free during the entire cosmic evolution of the universe even at $H = 0$ in a bouncing scenario. The four- and five-parameter entropies shown in Eq.~(\ref{intro-1}) and in Eq.~(\ref{intro-4}), respectively, are regarded as the minimal construction(s) for the generalized version of entropy depending on the fact that whether the evolutionary phases of the universe pass through $H = 0$ or not.

Being the minimal construction, $S_\mathrm{4}$ and $S_\mathrm{5}$ have been explored in investigating various cosmic phenomena in the context of entropic cosmology where the presence of entropy generates an effective energy density that hints for inflation, dark energy, or even, a bounce \cite{Nojiri:2022aof,Odintsov:2022qnn,Odintsov:2023qfj,Odintsov:2023vpj}. In particular, the entropic cosmology based on the four-parameter $S_\mathrm{4}$ proves to be viable to describe from inflation to reheating era of the early universe \cite{Odintsov:2023vpj}. Actually, the presence of the entropic parameters in the $S_\mathrm{4}$ ensures a continuous evolution of the Hubble parameter from a quasi de-Sitter phase during the inflation to a power law phase during the reheating stage dominated by a constant EoS parameter. Moreover, owing to the non-singular character, the occurrence of the five-parameter $S_\mathrm{5}$ in entropic cosmology triggers a stable bouncing universe that is also consistent with the cosmological perturbation and the associated observational data \cite{Odintsov:2022qnn,Odintsov:2023qfj}.

Such interest in generalized entropies compels us to understand its microscopic origin. This understanding is also important from the perspective that being a thermodynamic quantity associated with the apparent horizon, the generalized entropy needs to have a microscopic root under the ensemble average. In the present work, we will address the possible microscopic origin of the generalized entropies, in particular, we consider how various entropies of the form $S_\mathrm{3}$, $S_\mathrm{4}$, $S_\mathrm{5}$ and $S_\mathrm{6}$ can appear from canonical and grand-canonical description of some thermodynamic system. We should mention that some of our authors showed a microscopic root of the generalized entropies through microcanonical and canonical descriptions, however from a different point of view \cite{Nojiri:2023ikl}. Note that, in our present analysis, we include the grand-canonical description and also consider the thermodynamic systems with {\it non-zero} potential, which makes the present scenario essentially different from earlier ones. The consideration of a general potential includes the gravitating system interacting with some suitable gravitational potential and thus could be helpful to clarify the structure of quantum gravity.


\section{Canonical description}\label{sec-c}
For the microscopic origin of various generalized entropies, let us start from their Taylor series expansion as,
\begin{eqnarray}
 S_\mathrm{3}\left(\alpha,\delta,\gamma\right)&=&\sum_{n=0}^{\infty} \frac{f_\mathrm{3}\left(\alpha,\delta,\gamma\right)}{n!}S^{n}\, ,\nonumber\\
 S_\mathrm{4}\left(\alpha_{\pm},\delta,\gamma\right)&=&\sum_{n=0}^{\infty} \frac{f_\mathrm{4}\left(\alpha_{\pm},\delta,\gamma\right)}{n!}S^{n}\, ,\nonumber\\
 S_\mathrm{5}\left(\alpha_{\pm},\delta,\gamma,\epsilon\right)&=&\sum_{n=0}^{\infty} \frac{f_\mathrm{5}\left(\alpha_{\pm},\delta,\gamma,\epsilon\right)}{n!}S^{n}\, ,\nonumber\\
 S_\mathrm{6}\left(\alpha_{\pm},\delta_{\pm},\gamma_{\pm}\right)&=&\sum_{n=0}^{\infty} \frac{f_\mathrm{6}\left(\alpha_{\pm},\delta_{\pm},\gamma_{\pm}\right)}{n!}S^{n}\, ,\label{canonical-1}
\end{eqnarray}
where the functions $f_\mathrm{g} = \left\{f_\mathrm{3},f_\mathrm{4},f_\mathrm{5},f_\mathrm{6}\right\}$ are defined by $f_\mathrm{g} = \frac{d^nS_\mathrm{g}}{dS^n}\bigg|_{S=0}$ with $S_\mathrm{g} = \left\{S_\mathrm{3},S_\mathrm{4},S_\mathrm{5},S_\mathrm{6}\right\}$, in particular, they are obtained as follows:
\begin{eqnarray}
 f_\mathrm{3}\left(\alpha,\delta,\gamma\right)&=&\frac{\delta}{\gamma}\left[\left(\frac{\alpha_+}{\delta}\right)^n(\delta-1)(\delta-2).....(\delta-n+1)\right]~,\nonumber\\
 f_\mathrm{4}\left(\alpha_{\pm},\delta,\gamma\right)&=&\frac{\delta}{\gamma}\left[\left(\frac{\alpha_+}{\delta}\right)^n(\delta-1)(\delta-2).....(\delta-n+1) - (-1)^n\left(\frac{\alpha_-}{\delta}\right)^n(\delta+1)(\delta+2).....(\delta+n-1)\right]~,\nonumber\\
 f_\mathrm{5}\left(\alpha_{\pm},\delta,\gamma,\epsilon\right)&=&\frac{1}{\gamma}\left. \left[\frac{\partial^n}{\partial S^n}\left\{1 + \frac{1}{\epsilon}\tanh\left(\frac{\epsilon \alpha_+}{\delta}~S\right)\right\}^{\delta} - \frac{\partial^n}{\partial S^n}\left\{1 + \frac{1}{\epsilon}\tanh\left(\frac{\epsilon \alpha_-}{\delta}~S\right)\right\}^{-\delta}\right]\right|_{S=0}
 \label{canonical-2a}
 \end{eqnarray}
 and
\begin{eqnarray}
 f_\mathrm{6}\left(\alpha_{\pm},\delta_{\pm},\gamma_{\pm}\right) = \frac{1}{\alpha_+ + \alpha_-}&\bigg[&\left(\frac{\alpha_+}{\delta_+}\right)^n\delta_+(\delta_+ - 1)(\delta_+ - 2).....(\delta_+ - n+1)\nonumber\\
 &-&(-1)^n\left(\frac{\alpha_-}{\delta_-}\right)^n\delta_-(\delta_- + 1)(\delta_- + 2).....(\delta_- + n-1)\bigg]
 \label{canonical-2}
\end{eqnarray}
respectively. Such Taylor series expansions are valid due to the analytic behaviour of $S_\mathrm{g} = \left\{S_\mathrm{3},S_\mathrm{4},S_\mathrm{5},S_\mathrm{6}\right\}$ with respect to the Bekenstein-Hawking entropy variable $S$.

In canonical prescription, the phase space density of a thermodynamic system having $N$ number of particles is given by,
\begin{eqnarray}
 \rho_\mathrm{c}\left(q_j,p_j\right) = \frac{\mathrm{exp}\left(-\beta H\right)}{Z(T,V,N)}\, ,
 \label{canonical-3}
\end{eqnarray}
where $\beta = 1/\left(kT\right)$ (with $k$ being the Boltzmann constant and $T$ is the temperature of the system), $j$ runs from $j=1$ to $j=3N$, $\left\{q_j,p_j\right\}$ represent the generalized coordinate and generalized momenta of the system, respectively. Moreover, $H(q_j,p_j)$ symbolizes the Hamiltonian that indeed contains the potential of the system (if any), and
\begin{eqnarray}
 Z(T,V,N) = \int \frac{d^{3N}q~d^{3N}p}{h^{3N}}e^{-\beta H}\, ,
 \label{canonical-4}
\end{eqnarray}
is known as the partition function which is a macroscopic quantity and depends on temperature ($T$), volume ($V$) and number of particles ($N$) of the system. The possible microstates, in this case, lie within $H(q_j,p_j) = \left\langle H \right\rangle \pm \Delta H$ in [6N] dimensional phase space volume, where $\left\langle H \right\rangle$ is the ensemble average of $H$. The above form of $Z(T,V,N)$ ensures that the total probability for finding the system in any of the possible microstates is unity, i.e.
\begin{eqnarray}
 \int \frac{d^{3N}q~d^{3N}p}{h^{3N}} \rho_\mathrm{c}(q_j,p_j) = 1\, .
 \label{canonical-5}
\end{eqnarray}
Consequently the ensemble average of a general microscopic quantity $v(q_j,p_j)$ is defined by,
\begin{eqnarray}
 \left\langle v(q,p) \right\rangle = \int \frac{d^{3N}q~d^{3N}p}{h^{3N}} v(q,p)\rho_\mathrm{c}(q,p)\, .
 \label{canonical-6}
\end{eqnarray}
The Gibbs entropy which we now denote by $S_\mathrm{0}$, is given by the ensemble average of the microscopic quantity namely $\left\langle -k\ln{\rho_\mathrm{c}}\right\rangle$:
\begin{eqnarray}
 S_\mathrm{0} = \left\langle -k\ln{\rho_\mathrm{c}} \right\rangle&=&-k\int \frac{d^{3N}q~d^{3N}p}{h^{3N}} \rho_\mathrm{c}\ln{\rho_\mathrm{c}}\nonumber\\
 &=&k\beta\left\langle H \right\rangle + k\ln{Z}\, .
 \label{canonical-7}
\end{eqnarray}
where, to derive the last equality, we use Eq.~(\ref{canonical-5}) and Eq.~(\ref{canonical-6}). The expression $S_\mathrm{0} = \left\langle -k\ln{\rho_\mathrm{c}}\right\rangle$ actually defines the macroscopic quantity like the entropy from the microscopic point of view. Consequently the ensemble average of $(-k\ln{\rho_\mathrm{c}})^2$ turns out to be,
\begin{eqnarray}
 \left\langle \left(-k\ln{\rho_\mathrm{c}}\right)^2 \right\rangle&=&k^2 \int \frac{d^{3N}q~d^{3N}p}{h^{3N}} \rho_\mathrm{c}\left(\ln{\rho_\mathrm{c}}\right)^2\nonumber\\
 &=&k^2\beta^2\left\langle H^2 \right\rangle + 2k^2\beta \left\langle H \right\rangle\ln{Z} + \left(k\ln{Z}\right)^2\, ,
 \label{canonical-8}
\end{eqnarray}
which can be equivalently expressed by,
\begin{eqnarray}
 \left\langle \left(-k\ln{\rho_\mathrm{c}}\right)^2 \right\rangle = S_\mathrm{0}^2 + k^2\beta^2\sigma_\mathrm{2}(H)\, .
 \label{canonical-9}
\end{eqnarray}
Here $\sigma_\mathrm{2}(H) = \left\langle H^2 \right\rangle - \left\langle H \right\rangle^2$ represents the deviation of energy from its mean value at quadratic order. Therefore from Eq.~(\ref{canonical-9}), we have the microscopic interpretation of $S_\mathrm{0}^2$ as follows,
\begin{eqnarray}
 S_\mathrm{0}^2 = \left\langle \left(-k\ln{\rho_\mathrm{c}}\right)^2\right\rangle - k^2\beta^2\sigma_\mathrm{2}(H)\, ,
 \label{canonical-10}
\end{eqnarray}
i.e., $S_\mathrm{0}^2$ is the ensemble average of $\left(-k\ln{\rho_\mathrm{c}}\right)^2$ with the factor containing $\sigma_\mathrm{2}(H)$. For simple additive thermodynamic system, $\frac{\sigma_\mathrm{2}(H)}{\left\langle H \right\rangle}$ gets proportional to $\frac{1}{\sqrt{N}}$ which tends to zero in the thermodynamic limit ($N \rightarrow \infty$), and thus Eq.~(\ref{canonical-10}) reduces to $S_\mathrm{0}^2 = \left\langle\left(-k\ln{\rho_\mathrm{c}}\right)^2\right\rangle$. However for systems that do not possess such behaviour, for instance, the non-additive system, $\frac{\sigma_\mathrm{2}(H)}{\left\langle H \right\rangle}$ does not vanish even in the thermodynamic limit and consequently $S_\mathrm{0}^2$ is given by Eq.~(\ref{canonical-10}). Thus, in general, we will continue with Eq.~(\ref{canonical-10}) for the expression of $S_\mathrm{0}^2$. Similarly the ensemble average of $\left(-k\ln{\rho_\mathrm{c}}\right)^3$ is given by,
\begin{eqnarray}
 \left\langle \left(-k\ln{\rho_\mathrm{c}}\right)^3\right\rangle&=&-k^3 \int \frac{d^{3N}q~d^{3N}p}{h^{3N}} \rho_\mathrm{c}\left(\ln{\rho_\mathrm{c}}\right)^3\nonumber\\
 &=&k^3\beta^3\left\langle H^3 \right\rangle + 3k^3\beta^2\left\langle H^2 \right\rangle \ln{Z} + 3k^2\beta\left\langle H \right\rangle \ln{Z} + \left(k\ln{Z}\right)^3\, ,
 \label{canonical-11}
\end{eqnarray}
which immediately leads to $S_\mathrm{0}^3$ as,
\begin{eqnarray}
 S_\mathrm{0}^3 = \left\langle \left(-k\ln{\rho_\mathrm{c}}\right)^3\right\rangle - k^3\beta^3\sigma_\mathrm{3}(H) - 3k^3\beta^2\sigma_\mathrm{2}(H)\ln{Z}\, ,
 \label{canonical-12}
\end{eqnarray}
with $\sigma_\mathrm{3}(H) = \left\langle H^3 \right\rangle - \left\langle H \right\rangle^3$. The analogy of Eq.~(\ref{canonical-10}) and Eq.~(\ref{canonical-12}) yields $S_\mathrm{0}^n$ as ($n$ being a positive integer),
\begin{eqnarray}
 S_\mathrm{0}^n = \left\langle \left(-k\ln{\rho_\mathrm{c}}\right)^n\right\rangle - \sum_{i=2}^{n}\frac{n!}{i!(n-i)!}\left(k\beta\right)^{i}\sigma_i(H)\left(k\ln{Z}\right)^{n-i}\, ,
 \label{canonical-13}
\end{eqnarray}
where $\sigma_i(H) = \left\langle H^i \right\rangle - \left\langle H \right\rangle^i$ represents the statistical fluctuation of energy from its ensemble average at $i$-th order. It may be noted that the lower limit of the sum in the right-hand side of Eq.~(\ref{canonical-13}) can be safely taken from $i = 1$ due to the fact that $\sigma_\mathrm{1}(H) = 0$. By using
\begin{eqnarray}
 \left\langle H^i \right\rangle = \frac{1}{Z} \int \frac{d^{3N}q~d^{3N}p}{h^{3N}} e^{-\beta H} H^{i}\, ,
 \label{canonical-14}
\end{eqnarray}
one can express $\sigma_i(H)$, in terms of the partition function, and is given by,
\begin{eqnarray}
 \sigma_i(H) = \left(-1\right)^{i}\left\{\frac{1}{Z}\frac{\partial^{i}Z}{\partial\beta^{i}} - \left(\frac{1}{Z}\frac{\partial Z}{\partial\beta}\right)^i\right\}\, .
 \label{canonical-15}
\end{eqnarray}
Owing to the expression of $S_\mathrm{0}^n$ in Eq.~(\ref{canonical-15}), we define an entropy similar to the form of generalized entropy ($S_\mathrm{g} = \left\{S_\mathrm{3},S_\mathrm{4},S_\mathrm{5},S_\mathrm{6}\right\}$), as follows:
\begin{eqnarray}
 S_\mathrm{can}&=&\sum_{n=0}^{\infty} \frac{f_\mathrm{g}\left(\alpha,\delta,\gamma,.....\right)}{n!}S_\mathrm{0}^n\nonumber\\
 &=&\sum_{n=0}^{\infty} \frac{f_\mathrm{g}\left(\alpha,\delta,\gamma,.....\right)}{n!}\left\{\left\langle \left(-k\ln{\rho_\mathrm{c}}\right)^n\right\rangle - \sum_\mathrm{i=2}^{n}\frac{n!}{i!(n-i)!}\left(k\beta\right)^{i}\sigma_i(H)\left(k\ln{Z}\right)^{n-i}\right\}\, ,
 \label{canonical-16}
\end{eqnarray}
where $f_\mathrm{g} = \left\{f_\mathrm{3},f_\mathrm{4},f_\mathrm{5},f_\mathrm{6}\right\}$ are shown in Eq.~(\ref{canonical-2a}) (also see Eq.~(\ref{canonical-2})) and $\sigma_i(H)$ is obtained in Eq.~(\ref{canonical-15}). The comparison of Eq.~(\ref{canonical-1}) and Eq.~(\ref{canonical-16}) clearly argues that $S_\mathrm{can}$ has a similar form of various generalized entropies depending on the form of $f_\mathrm{g}$. For instance --- with $f_\mathrm{g} = f_\mathrm{3}\left(\alpha,\delta,\gamma\right)$ (see Eq.~(\ref{canonical-2a})), the $S_\mathrm{can}$ looks like the 3-parameter generalized entropy $S_\mathrm{3}$ of Eq.~(\ref{intro-3}), while for $f_\mathrm{g} = f_\mathrm{4}\left(\alpha_{\pm},\delta,\gamma\right)$ (see Eq.~(\ref{canonical-2a})), the $S_\mathrm{can}$ becomes of the form of the 4-parameter generalized $S_\mathrm{4}$ of Eq.~(\ref{intro-1}) etc. In particular,
\begin{eqnarray}
 S_\mathrm{can}&=&\sum_{n=0}^{\infty} \frac{f_\mathrm{3}\left(\alpha,\delta,\gamma\right)}{n!}\left\{\left\langle \left(-k\ln{\rho_\mathrm{c}}\right)^n\right\rangle - \sum_\mathrm{i=2}^{n}\frac{n!}{i!(n-i)!}\left(k\beta\right)^{i}\sigma_i(H)\left(k\ln{Z}\right)^{n-i}\right\}\nonumber\\
&=&\frac{1}{\gamma}\left[\left(1 + \frac{\alpha}{\delta}~S_\mathrm{0}\right)^{\delta} - 1\right]\, ,
\label{canonical-17}
\end{eqnarray}
and
\begin{eqnarray}
 S_\mathrm{can}&=&\sum_{n=0}^{\infty} \frac{f_\mathrm{4}\left(\alpha_{\pm},\delta,\gamma\right)}{n!}\left\{\left\langle \left(-k\ln{\rho_\mathrm{c}}\right)^n\right\rangle - \sum_\mathrm{i=2}^{n}\frac{n!}{i!(n-i)!}\left(k\beta\right)^{i}\sigma_i(H)\left(k\ln{Z}\right)^{n-i}\right\}\nonumber\\
&=&\frac{1}{\gamma}\left[\left(1 + \frac{\alpha_+}{\delta}~S_\mathrm{0}\right)^{\delta}
 - \left(1 + \frac{\alpha_-}{\delta}~S_\mathrm{0}\right)^{-\delta}\right]\, ,
\label{canonical-18}
\end{eqnarray}
respectively, with recall that $S_\mathrm{0} = \left\langle -k\ln{\rho_\mathrm{c}}\right\rangle$. The convergence of the above series in Eq.~(\ref{canonical-17}) (or in Eq.~(\ref{canonical-18})) is indeed ensured by Eq.~(\ref{canonical-1}). In this way, by considering $f_\mathrm{g} = f_\mathrm{5}\left(\alpha_{\pm},\delta,\gamma,\epsilon\right)$ (see Eq.~(\ref{canonical-2a})) or $f_\mathrm{g} = f_\mathrm{6}\left(\alpha_{\pm},\delta_{\pm},\gamma_{\pm}\right)$ (see Eq.~(\ref{canonical-2})), the $S_\mathrm{can}$ will be like the five-parameter or six-parameter generalized entropy, respectively. At this stage, it deserves mentioning that due to the consideration of potential in the Hamiltonian, the present formalism for microscopic interpretation of generalized entropy also includes the gravitating system (in general) where, for example, the particles' interact with Newtonian potential given by
\begin{eqnarray}
 V(q_j) = \sum_{k,l=1;k\neq l}^{N}-\frac{Gm^2}{\left|\vec{r}_k - \vec{r}_l\right|}\, ,\nonumber
\end{eqnarray}
where $m$ and $\vec{r}$ are the mass and position vector of the particles, respectively.\\

Thus $S_\mathrm{can} \equiv S_\mathrm{g} = \left\{S_\mathrm{3},S_\mathrm{4},S_\mathrm{5},S_\mathrm{6}\right\}$, and hence the generalized entropies in the canonical prescription can be interpreted as the statistical ensemble average of a series of microscopic quantity(ies) given by various powers of $\left(-k\ln{\rho_\mathrm{c}}\right)^n$ (with $n$ is a positive integer), along with $\sigma_i(H)$ representing the fluctuation of the Hamiltonian. Moreover, the coefficients in the series fix the nature of the generalized entropy namely $S_\mathrm{3}$ or $S_\mathrm{4}$ or $S_\mathrm{5}$ or $S_\mathrm{6}$, respectively.

\section{Grand-Canonical description}\label{sec-gc}
The grand-canonical phase space density, for a thermodynamic system having Hamiltonian $H$ and chemical potential $\mu$, is defined by,
\begin{eqnarray}
 \rho_\mathrm{gc}\left(q_j,p_j,N\right) = \frac{\mathrm{exp}\left\{-\beta\left(H-\mu N\right)\right\}}{\mathcal{Z}(T,V,\mu)}\, ,
 \label{g-canonical-1}
\end{eqnarray}
where, once again, $j$ runs from $j=1$ to $j=3N$ in [6N] dimensional phase space. Along with the Hamiltonian, the particle number in grand-canonical case also fluctuates around its statistical mean value $\left\langle N \right\rangle$; and thus, the possible microstates lie within $H = \left\langle H \right\rangle \pm \Delta H$ and $N = \left\langle N \right\rangle \pm \Delta N$ in phase space volume (where $\left\langle N \right\rangle$ being the ensemble average of $N$). Thus a single microstate is characterized by $\left\{q_j,p_j,N\right\}$ in a grand-canonical ensemble, unlike the canonical ensemble where the particle number is fixed to $\left\langle N \right\rangle$ and the microstates are designated by $\left\{q_j,p_j\right\}$. Moreover, $\mathcal{Z}$ is known as a grand-canonical partition function with the following form,
\begin{eqnarray}
 \mathcal{Z}(T,V,\mu) = \sum_\mathrm{N} \int \frac{d^{3N}q~d^{3N}p}{h^{3N}}~e^{-\beta\left(H-\mu N\right)}\, .
 \label{g-canonical-2}
\end{eqnarray}
The form of $\mathcal{Z}$ ensures the conservation of probability over all the possible microstates. Consequently the ensemble average of a microscopic quantity $v(q_j,p_j,N)$ in grand-canonical description is given by,
\begin{eqnarray}
 \left\langle v(q,p,N) \right\rangle = \sum_\mathrm{N} \int \frac{d^{3N}q~d^{3N}p}{h^{3N}} v(q,p,N)\rho_\mathrm{gc}(q,p,N)\, ,
 \label{g-canonical-3}
\end{eqnarray}
where $\rho_\mathrm{gc}$ is shown in Eq.~(\ref{g-canonical-1}).

For grand-canonical description, the Gibbs entropy symbolized by $S_\mathrm{0}$ is given by the ensemble average of $\ln{\rho_\mathrm{gc}}$, in particular,
\begin{eqnarray}
 S_\mathrm{0} = \left\langle -k\ln{\rho_\mathrm{gc}} \right\rangle&=&-k\sum_\mathrm{N} \int \frac{d^{3N}q~d^{3N}p}{h^{3N}} \rho_\mathrm{gc}\ln{\rho_\mathrm{gc}}\nonumber\\
 &=&k\beta\left\langle H \right\rangle - k\mu\left\langle N \right\rangle + k\ln{\mathcal{Z}}\, ,
 \label{g-canonical-4}
\end{eqnarray}
where we use Eq.~(\ref{g-canonical-3}) and the conservation of phase space density. Consequently the ensemble average of $\left(-k\ln{\rho_\mathrm{gc}}\right)^2$ turns out to be,
\begin{eqnarray}
 \left\langle \left(-k\ln{\rho_\mathrm{gc}}\right)^2\right\rangle&=&k^2 \sum_\mathrm{N} \int \frac{d^{3N}q~d^{3N}p}{h^{3N}} \rho_\mathrm{gc}\left(\ln{\rho_\mathrm{gc}}\right)^2\nonumber\\
 &=&k^2\beta^2\left\langle\left(H - \mu N\right)^2 \right\rangle + 2k^2\beta\left\langle H-\mu N \right\rangle\ln{\mathcal{Z}} + \left(k\ln{\mathcal{Z}}\right)^2\, ,
 \label{g-canonical-5}
\end{eqnarray}
which, by using Eq.~(\ref{g-canonical-4}), can be written as,
\begin{eqnarray}
 \left\langle \left(-k\ln{\rho_\mathrm{gc}}\right)^2\right\rangle = S_\mathrm{0}^2 + k^2\beta^2\sigma_\mathrm{2}(H-\mu N)\, ,
 \label{g-canonical-6}
\end{eqnarray}
with $\sigma_\mathrm{2}(H - \mu N) = \left\langle (H - \mu N)^2\right\rangle - \left\langle H - \mu N\right\rangle^2$. The above equation immediately leads to the microscopic interpretation of $S_\mathrm{0}^2$ as follows:
\begin{eqnarray}
 S_\mathrm{0}^2 = \left\langle \left(-k\ln{\rho_\mathrm{gc}}\right)^2\right\rangle - k^2\beta^2\sigma_\mathrm{2}(H-\mu N)\, ,
 \label{g-canonical-7}
\end{eqnarray}
i.e., $S_\mathrm{0}^2$ is the ensemble average of $\left(-k\ln{\rho_\mathrm{gc}}\right)^2$ corrected by a factor containing $\sigma_\mathrm{2}(H-\mu N)$. The comparison of Eq.~(\ref{canonical-10}) and Eq.~(\ref{g-canonical-7}) clearly reflects the fact that $S_\mathrm{0}^2$ in the grand-canonical scenario contains the fluctuation of $N$ (around $\left\langle N \right\rangle$) through $\sigma_\mathrm{2}(H- \mu N)$, unlike to that of in the canonical case. Therefore the ensemble average of $\left(-k\ln{\rho_\mathrm{c}}\right)^3$ is given by,
\begin{eqnarray}
 \left\langle \left(-k\ln{\rho_\mathrm{gc}}\right)^3\right\rangle&=&-k^3 \sum_\mathrm{N} \int \frac{d^{3N}q~d^{3N}p}{h^{3N}} \rho_\mathrm{gc}\left(\ln{\rho_\mathrm{gc}}\right)^3\nonumber\\
 &=&k^3\beta^3\left\langle\left(H - \mu N\right)^3 \right\rangle + 3k^3\beta^2\left\langle\left(H-\mu N\right)^2 \right\rangle \ln{\mathcal{Z}} + 3k^2\beta\left\langle H-\mu N \right\rangle \ln{\mathcal{Z}} + \left(k\ln{\mathcal{Z}}\right)^3\, ,
 \label{g-canonical-8}
\end{eqnarray}
which yields $S_\mathrm{0}^3$ as,
\begin{eqnarray}
 S_\mathrm{0}^3 = \left\langle \left(-k\ln{\rho_\mathrm{gc}}\right)^3\right\rangle - k^3\beta^3\sigma_\mathrm{3}(H-\mu N) - 3k^3\beta^2\ln{\mathcal{Z}}\sigma_\mathrm{2}(H-\mu N)\, ,
 \label{g-canonical-9}
\end{eqnarray}
with $\sigma_\mathrm{3}(H - \mu N) = \left\langle (H - \mu N)^3\right\rangle - \left\langle H - \mu N\right\rangle^3$. Continuing as Eq.~(\ref{g-canonical-7}) and Eq.~(\ref{g-canonical-9}), we may write the expression of $S_\mathrm{0}^n$ (with $n$ being a positive integer) as follows,
\begin{eqnarray}
 S_\mathrm{0}^n = \left\langle \left(-k\ln{\rho_\mathrm{gc}}\right)^n\right\rangle - \sum_\mathrm{i=2}^{n}\frac{n!}{i!(n-i)!}\left(k\beta\right)^{i}\sigma_i(H-\mu N)\left(k\ln{\mathcal{Z}}\right)^{n-i}\, ,
 \label{g-canonical-10}
\end{eqnarray}
where $\sigma_i(H-\mu N)$ represents the deviation of $\left(H-\mu N\right)$ from its mean value at $i$-th order. Such deviation can be determined in terms of $\mathcal{Z}$ due to the following expression,
\begin{eqnarray}
 \left\langle\left(H - \mu N\right)^i \right\rangle = \frac{1}{\mathcal{Z}} \sum_\mathrm{N} \int \frac{d^{3N}q~d^{3N}p}{h^{3N}} e^{-\beta(H-\mu N)} \left(H - \mu N\right)^{i}\, ,
 \label{g-canonical-11}
\end{eqnarray}
which results in,
\begin{eqnarray}
 \sigma_i(H-\mu N) = \left(-1\right)^{i}\left\{\frac{1}{\mathcal{Z}}\frac{\partial^{i}\mathcal{Z}}{\partial\beta^{i}} - \left(\frac{1}{\mathcal{Z}}\frac{\partial \mathcal{Z}}{\partial\beta}\right)^i\right\}\, .
 \label{g-canonical-12}
\end{eqnarray}
By using Eq.~(\ref{g-canonical-10}), and similar to the canonical scenario, we define the following entropy in the grand-canonical description, similar to the form of generalized entropies, as
\begin{eqnarray}
 S_\textrm{gr-can} = \sum_{n=0}^{\infty} \frac{f_\mathrm{g}\left(\alpha,\delta,\gamma,.....\right)}{n!}\left\{\left\langle \left(-k\ln{\rho_\mathrm{gc}}\right)^n\right\rangle - \sum_\mathrm{i=2}^{n}\frac{n!}{i!(n-i)!}\left(k\beta\right)^{i}\sigma_i(H-\mu N)\left(k\ln{\mathcal{Z}}\right)^{n-i}\right\}\, ,
 \label{g-canonical-13}
\end{eqnarray}
where $f_\mathrm{g} = \left\{f_\mathrm{3},f_\mathrm{4},f_\mathrm{5},f_\mathrm{6}\right\}$ are shown in Eq.~(\ref{canonical-2a}) (and in Eq.~(\ref{canonical-2})). Clearly $S_\textrm{gr-can}$ has similar form of various generalized entropies $S_\mathrm{g} = \left\{S_\mathrm{3},S_\mathrm{4},S_\mathrm{5},S_\mathrm{6}\right\}$ depending on $f_\mathrm{g}$. With $f_\mathrm{g} = f_\mathrm{3}\left(\alpha,\delta,\gamma\right)$ of Eq.~(\ref{canonical-2a}),
\begin{eqnarray}
 S_\textrm{gr-can}&=&\sum_{n=0}^{\infty} \frac{f_\mathrm{3}\left(\alpha,\delta,\gamma\right)}{n!}\left\{\left\langle \left(-k\ln{\rho_\mathrm{gc}}\right)^n\right\rangle - \sum_\mathrm{i=2}^{n}\frac{n!}{i!(n-i)!}\left(k\beta\right)^{i}\sigma_i(H-\mu N)\left(k\ln{\mathcal{Z}}\right)^{n-i}\right\}\nonumber\\
&=&\frac{1}{\gamma}\left[\left(1 + \frac{\alpha}{\delta}~S_\mathrm{0}\right)^{\delta} - 1\right]\, ,
\label{g-canonical-14}
\end{eqnarray}
or, with $f_\mathrm{g} = f_\mathrm{4}\left(\alpha_{\pm},\delta,\gamma\right)$ of Eq.~(\ref{canonical-2a}),
\begin{eqnarray}
 S_\textrm{gr-can}&=&\sum_{n=0}^{\infty} \frac{f_\mathrm{4}\left(\alpha_{\pm},\delta,\gamma\right)}{n!}\left\{\left\langle \left(-k\ln{\rho_\mathrm{gc}}\right)^n\right\rangle - \sum_\mathrm{i=2}^{n}\frac{n!}{i!(n-i)!}\left(k\beta\right)^{i}\sigma_i(H-\mu N)\left(k\ln{\mathcal{Z}}\right)^{n-i}\right\}\nonumber\\
&=&\frac{1}{\gamma}\left[\left(1 + \frac{\alpha_+}{\delta}~S_\mathrm{0}\right)^{\delta}
 - \left(1 + \frac{\alpha_-}{\delta}~S_\mathrm{0}\right)^{-\delta}\right]\, ,
\label{g-canonical-15}
\end{eqnarray}
etc. Similarly $S_\textrm{gr-can}$ can be made of the same form as five-parameter or six-parameter generalized entropy by considering $f_\mathrm{g} = f_\mathrm{5}\left(\alpha_{\pm},\delta,\gamma,\epsilon\right)$ or $f_\mathrm{g} = f_\mathrm{6}\left(\alpha_{\pm},\delta_{\pm},\gamma_{\pm}\right)$, respectively.

Therefore in the grand-canonical scenario, $S_\textrm{gr-can} \equiv \left\{S_\mathrm{3},S_\mathrm{4},S_\mathrm{5},S_\mathrm{6}\right\}$ and thus the generalized entropies (depending on various parameters) can be microscopically interpreted as the ensemble average of a series of $\left(-k\ln{\rho_\mathrm{gc}}\right)^{n}$ (with $n=0,1,2...$ being a positive integer), along with $\sigma_i(H-\mu N)$ designating the statistical fluctuations of Hamiltonian and number of particles of the system under consideration. Depending on the coefficients in the series, $S_\textrm{gr-can}$ reduces to the form like $S_\mathrm{3}$ or $S_\mathrm{4}$ or $S_\mathrm{5}$ or $S_\mathrm{6}$. Furthermore by comparing Eq.~(\ref{canonical-18}) and Eq.~(\ref{g-canonical-15}), note that the microscopic description of generalized entropy(ies) in canonical and grand-canonical ensemble appear to be more-or-less same.\\

In the present context, it is important to note the presence of chemical potential ($\mu$) in the grand-canonical description. Chemical potential represents the work necessary to add a particle to the system by maintaining the equilibrium of the same. For maintaining the equilibrium, one can not simply add the particle at rest into the system, rather it has to have a certain energy that is comparable to the mean energy of all the other particles. The entropy of a certain thermodynamic system depends on the corresponding statistics, i.e., depending on the additive or non-additive statistics, the Gibbs entropy or the Tsallis/R\'{e}nyi entropy becomes applicable. However the generalized entropy is applicable to any thermodynamic system irrespective of its additive or non-additive nature, as the generalized entropy converges to various known entropies for suitable parameter values. In particular --- for additive system(s), the generalized entropy converges to the Gibbs entropy (or to some other extensive entropy) with certain parameter values, while for non-additive system(s), the generalized entropy goes to the Tsallis or to the R\'{e}nyi entropy (or to some other non-extensive entropy) with suitable parameter representatives. If the system under consideration is closed in nature, then it is microscopically described by canonical ensemble, and consequently, the microscopic interpretation of generalized entropy is given by Eq.~(\ref{canonical-16}). On other hand, an open system, where the particle number fluctuates, is described by grand-canonical ensemble and thus the generalized entropy is microscopically interpreted by Eq.~(\ref{g-canonical-13}). Regarding the grand-canonical ensemble, the respective phase space density explicitly depends on chemical potential which accounts the fluctuation of particle number of the system. Moreover in the case of canonical ensemble, the $S_\mathrm{0}$ in Eq.~(\ref{canonical-18}) (or in Eq.~(\ref{canonical-17})) is given by $\langle -k\ln{\rho_\mathrm{c}} \rangle$ which results to,
\begin{eqnarray}
 S_\mathrm{0} = k\beta\left\langle H \right\rangle + k\ln{Z}
 \nonumber
\end{eqnarray}
(with $Z$ being the canonical partition function, see Eq.~(\ref{canonical-7})); while in the grand-canonical description, the $S_\mathrm{0}$ in Eq.~(\ref{g-canonical-15}) (or in Eq.~(\ref{g-canonical-14})) is defined by $\langle -k\ln{\rho_\mathrm{gc}} \rangle$ that finally leads to,
\begin{eqnarray}
 S_\mathrm{0} = k\beta\left\langle H \right\rangle - k\mu\left\langle N \right\rangle + k\ln{\mathcal{Z}}
 \nonumber
\end{eqnarray}
(where $\mathcal{Z}$ is the grand-canonical partition function, see Eq.~(\ref{g-canonical-4})). Therefore the right hand sides of Eq.~(\ref{canonical-18}) and Eq.~(\ref{g-canonical-15}) are, in general, not same. The validity of canonical and grand-canonical description is based on whether the system under consideration is closed or open in nature. For instance, in the cosmological context of apparent horizon thermodynamics, the matter contents inside of the horizon changes with the universe's expansion due to the dynamical nature of the apparent horizon, and thus, the thermodynamics of the system inside the horizon needs to be described by grand-canonical ensemble.\\

Here it deserves mentioning that in the above analysis of Sec.~[\ref{sec-c}] and Sec.~[\ref{sec-gc}], the generalized entropy(ies) are expanded by Taylor series around $S_\mathrm{0} = 0$ which occurs at zero temperature (as there is only one microstate at $T = 0$). This becomes possible because $S_\mathrm{g} \equiv \left\{S_\mathrm{3},S_\mathrm{4},S_\mathrm{5},S_\mathrm{6}\right\}$ are all order differentiable at $S_\mathrm{0} = 0$. However this may not be the case for some other entropy functions; for instance --- (a) the 1-particle excitation related entropy given by \cite{Telali:2021jju},
\begin{eqnarray}
 S_\textrm{1-p}\left(\zeta,\lambda\right) = S_\mathrm{0} + \zeta S_\mathrm{0}^{\lambda}~~,
 \label{new-1}
\end{eqnarray}
or, (b) the entropy with varying exponent having the form \cite{Nojiri:2019skr}
\begin{eqnarray}
 S_\mathrm{var}\left(\xi\right) = \xi S_\mathrm{0}^{\Omega(S_\mathrm{0})}~~.
 \label{new-2}
\end{eqnarray}
Here $\zeta$, $\xi$ and $\lambda$ are constants; while $\Omega(S_\mathrm{0})$ varies w.r.t. temperature (i.e. with energy scale), or equivalently, w.r.t. $S_\mathrm{0}$. Owing to the fact that both the exponents $\lambda$ and $\Omega(S_\mathrm{0})$ can lie within $0$ and $1$, the entropies $S_\textrm{1-p}$ and $S_\mathrm{var}$ are not all order differentiable at $S_\mathrm{0}=0$. Therefore in such cases, we need to expand the Taylor series of $S_\textrm{1-p}$ and $S_\mathrm{var}$ around some non-zero temperature, or equivalently, around $S_\mathrm{0} \neq 0$ (say at $S_\mathrm{0} = S_\mathrm{b}$). Consequently we may write the Taylor series as,
\begin{eqnarray}
 S_\textrm{1-p}\left(\zeta,\lambda\right)&=&\sum_{n=0}^{\infty} \frac{g\left(\zeta,\lambda\right)}{n!}\left(S_\mathrm{0} - S_\mathrm{b}\right)^{n}\, ,\nonumber\\
 S_\mathrm{var}\left(\xi\right)&=&\sum_{n=0}^{\infty} \frac{h\left(\xi\right)}{n!}\left(S_\mathrm{0} - S_\mathrm{b}\right)^{n}\, ,
 \label{new-3}
\end{eqnarray}
where $g\left(\zeta,\lambda\right)$ and $h\left(\xi\right)$ have the following forms:
\begin{align}
g\left(\zeta,\lambda\right) = \frac{d^nS_\textrm{1-p}}{dS_\mathrm{0}^n}\bigg|_{S_\mathrm{0} = S_\mathrm{b}}
\begin{cases}
= S_\mathrm{b} + \zeta S_\mathrm{b}^{\lambda}~;~~~~\textrm{for $n=0$}, & \\
= 1 + \lambda \zeta S_\mathrm{b}^{\lambda - 1}~;~~~~\textrm{for $n=1$}, & \\
= \zeta \lambda(\lambda-1)(\lambda-2)\cdots(\lambda-n+1)S_\mathrm{b}^{\lambda-n}~;~~~~\textrm{for $n\geq2$}
\end{cases}
\label{new-4}
\end{align}
and
\begin{align}
h\left(\xi\right) = \frac{d^nS_\mathrm{var}}{dS_\mathrm{0}^n}\bigg|_{S_\mathrm{0} = S_\mathrm{b}}
\begin{cases}
= \xi S_\mathrm{b}^{\Omega(S_\mathrm{b})}~;~~~~\textrm{for $n=0$}, & \\
= \xi \left[\Omega S_\mathrm{b}^{\Omega - 1} + \Omega'~S_\mathrm{b}^{\Omega}\ln{S_\mathrm{b}}\right]\bigg|_{S_\mathrm{b}}~;~~~~\textrm{for $n=1$}, & \\
= \textrm{and so on},
\end{cases}
\label{new-5}
\end{align}
respectively. The Taylor series of $S_\textrm{1-p}$ along with the microscopic interpretation of $S_\mathrm{0}^{n}$ in canonical and grand-canonical ensemble from Eq.~(\ref{canonical-13}) and Eq.~(\ref{g-canonical-10}) (respectively) allow us to express the $S_\textrm{1-p}$ in terms of phase space density as follows:
\begin{eqnarray}
 S_\textrm{1-p}&=&\sum_{n=0}^{\infty} \frac{g\left(\zeta,\lambda\right)}{n!}\left[\sum_{l=0}^{n}\frac{\left(-1\right)^{n-l}n!}{l!(n-l)!}S_\mathrm{b}^{n-l}\left\{\left\langle \left(-k\ln{\rho_\mathrm{c}}\right)^l\right\rangle - \sum_\mathrm{i=2}^{l}\frac{l!}{i!(l-i)!}\left(k\beta\right)^{i}\sigma_i(H)\left(k\ln{Z}\right)^{l-i}\right\}\right]
\label{new-6}
\end{eqnarray}
in canonical ensemble, and
\begin{eqnarray}
 S_\textrm{1-p}&=&\sum_{n=0}^{\infty} \frac{g\left(\zeta,\lambda\right)}{n!}\left[\sum_{l=0}^{n}\frac{\left(-1\right)^{n-l}n!}{l!(n-l)!}S_\mathrm{b}^{n-l}\left\{\left\langle \left(-k\ln{\rho_\mathrm{gc}}\right)^l\right\rangle - \sum_\mathrm{i=2}^{l}\frac{l!}{i!(l-i)!}\left(k\beta\right)^{i}\sigma_i(H- \mu N)\left(k\ln{\mathcal{Z}}\right)^{l-i}\right\}\right]
\label{new-7}
\end{eqnarray}
in grand-canonical ensemble. To arrive at the above expressions, we have used $\left(S_\mathrm{0} - S_\mathrm{b}\right)^{n} = \sum_{l=0}^{n}\frac{\left(-1\right)^{n-l}n!}{l!(n-l)!}S_\mathrm{b}^{n-l}S_\mathrm{0}^{l}$. The entropy with varying exponent, i.e. $S_\mathrm{var}$, can be similarly expressed by phase space density with the help of the respective Taylor series from Eq.(\ref{new-3}). Therefore, similar to $S_\mathrm{g} \equiv \left\{S_\mathrm{3},S_\mathrm{4},S_\mathrm{5},S_\mathrm{6}\right\}$, both the $S_\textrm{1-p}$ and $S_\mathrm{var}$ can be microscopically interpreted as the ensemble average of a series of $\left(-k\ln{\rho}\right)^{n}$ (with $n=0,1,2...$ being a positive integer and $\rho$ being the phase space density) along with a term representing the fluctuation of Hamiltonian and number of particles of the system under consideration (in case of canonical ensemble, the fluctuation on the particle number vanishes). However the only difference of $S_\textrm{1-p}$ and $S_\mathrm{var}$ with the generalized entropies is that the $S_\textrm{1-p}$ (and the $S_\mathrm{var}$) are not differentiable at $S_\mathrm{0} =0$ and thus they are expanded around non-zero temperature, unlike to the $S_\mathrm{g} \equiv \left\{S_\mathrm{3},S_\mathrm{4},S_\mathrm{5},S_\mathrm{6}\right\}$ which is all order differentiable at $S_\mathrm{0} = 0$ and expanded around the zero temperature.\\

Before concluding, it may be mentioned that we have used the semi-classical expression of phase space density, where the partition function is defined by an integral over $\frac{d^{3N}q~d^{3N}p}{h^{3N}}$, to arrive at the expressions of $S_\mathrm{can}$ (and $S_\textrm{gr-can}$) in Eq.~(\ref{canonical-18}) (and Eq.~(\ref{g-canonical-15})). However in the quantum realm, although the expression of $S_\mathrm{can}$ and $S_\textrm{gr-can}$ remain the same, the phase space density is replaced by a hermitian operator (known as density operator) given by,
\begin{eqnarray}
\hat{\rho}_\mathrm{c}&=&\frac{e^{-\beta \hat{H}}}{\mathrm{Tr}\left[e^{-\beta\hat{H}}\right]}\, ,~~~~~~~~~\mbox{for canonical ensemble}~;\nonumber\\
\hat{\rho}_\mathrm{gc}&=&\frac{e^{-\beta\left(\hat{H} - \mu \hat{N}\right)}}{\mathrm{Tr}\left[e^{-\beta\left(\hat{H} - \mu \hat{N}\right)}\right]}\, ,~~~~~~~~~\mbox{for grand canonical ensemble}~,\label{c-1}
\end{eqnarray}
respectively, where $\hat{H}$ and $\hat{N}$ are the Hamiltonian and particle number operator, respectively. Here the trace is taken over the microstates which, generally, are considered to be the common eigenstates of $\left\{\hat{H},\hat{N}\right\}$. Consequently, the ensemble average of $\left(-k\ln{\hat{\rho}}\right)$ is defined by,
\begin{eqnarray}
 \left\langle -k\ln{\hat{\rho}}\right\rangle = \mathrm{Tr}\left[\hat{\rho}\left(-k\ln{\hat{\rho}}\right)\right]\, .
 \label{c-2}
\end{eqnarray}
By using the above expressions, one can arrive at the same expressions of $S_\mathrm{can}$ (or $S_\textrm{gr-can}$) as obtained in Eq.~(\ref{canonical-18}) (or in Eq.~(\ref{g-canonical-15})) even in the quantum scenario.

\section{Summary}

We provide a possible microscopic origin of generalized entropy(ies) depending on a few parameters, which has been recently proposed as the most general entropy construction, that leads to all the known and apparently different entropies (like the Tsallis, the R\'{e}nyi, the Barrow, the Kaniadakis, the Sharma-Mittal and the loop quantum gravity entropy) as particular representatives. Our proposal is based on canonical and grand-canonical thermodynamic description. In particular, we show that in both the canonical and grand-canonical descriptions, the generalized entropies can be interpreted as the statistical ensemble average of a series of microscopic quantity(ies) given by various powers of $\left(-k\ln{\rho}\right)^n$ (with $n$ being a positive integer and $\rho$ symbolizes the phase space density of the respective ensemble), along with a term representing the fluctuation of Hamiltonian and number of particles of the system under consideration (in case of canonical ensemble, the fluctuation on the particle number vanishes). Moreover, the coefficients in the series actually fix the nature of generalized entropy namely the three, four, five or six-parameter dependent generalized entropy. Such microscopic interpretation holds even when the Hamiltonian contains a non-zero potential of the system under consideration. The presence of a general potential includes the gravitating system, as well, as interacting with some suitable gravitational potential, and thus the microscopic origin of generalized entropy could be helpful to clarify the structure of quantum gravity.

\end{document}